# TITLE PAGE

**Title:**

Minding rights: Mapping ethical and legal foundations of 'neurorights'

**Authors:**

*Sjors Ligthart, Marcello Ienca, Gerben Meynen, Fruzsina Molnar-Gabor, Roberto Andorno, Christoph Bublitz, Paul Catley, Lisa Claydon, Thomas Douglas, Nita Farahany, Joseph J. Fins, Sara Goering, Pim Haselager, Fabrice Jotterand, Andrea Lavazza, Allan McCay, Abel Wajnerman Paz, Stephen Rainey, Jesper Ryberg, Philipp Kellmeyer*[*]

**Affiliations:**

SL: Willem Pompe Institute for Criminal Law and Criminology, Utrecht University & Department of Criminal Law, Tilburg University, The Netherlands

MI: School of Medicine, Technical University of Munich (TUM), Germany & College of Humanities, Swiss Federal Institute of Technology in Lausanne (EPFL), Switzerland

GM: Willem Pompe Institute for Criminal Law and Criminology, Utrecht University & Department of Philosophy, Faculty of Humanities, Vrije Universiteit Amsterdam, Amsterdam, The Netherlands

FMG: Heidelberg University, Heidelberg, Germany

RA: University of Zurich, Switzerland

CB: Faculty of Law, Universität Hamburg, Germany

PC: School of Law, The Open University, UK

LC: School of Law, The Open University, UK

TD: Oxford University, UK

NF: Duke University, USA

JJF: Weill Cornell Medical College, New York, USA

PH: Donders Institute for Brain, Cognition, and Behaviour. Radboud University, Nijmegen, The Netherlands

FJ: Center for Bioethics and Medical Humanities, Medical College of Wisconsin, Milwaukee, WI, USA

AL: Centro Universitario Internazionale, Arezzo, Italy - https://orcid.org/0000-0003-2608-2609

AM: The University of Sydney Law School, Sydney, Australia

AWP: Instituto de Éticas Aplicadas, Pontificia Universidad Católica de Chile

SR: Ethics and Philosophy of Technology Section, Delft University, Delft, The Netherlands, https://orcid.org/0000-0002-5540-6046

JR: Roskilde University, Denmark.

PK: Department of Neurosurgery, University of Freiburg - Medical Center, c/o FRIAS, Alberstr. 19, D-79100 Freiburg im Breisgau, Germany

E-Mail: philipp.kellmeyer@uniklinik-freiburg.de

* corresponding author

**1. Introduction**

Advances in neurotechnology and artificial intelligence (AI) are challenging traditional boundaries of our brains and mental lives. Academic analyses of the ethical and legal implications of these advances in neurotechnology have now also spawned several initiatives at the national and international policymaking and lawmaking level.

In the scholarly literature, discussions of rights to our brains and mental experiences in human rights mainly focus on the legal protection of mental integrity, mental privacy and cognitive liberty by established human rights instruments, such as the Universal Declaration of Human Rights (UDHR), the International Covenant on Civil and Political Rights (ICCPR), the American Convention on Human Rights (ACHR), the European Convention on Human Rights (ECHR), and the Charter of Fundamental Rights of the European Union (CFR).[1] Meanwhile, The United Nations,[2] the Inter-American Juridical Committee,[3] the Committee on Bioethics of the Council of Europe,[4] UNSECO,[5] as well as the Organisation for Economic Co-operation and Development (OECD),[6] have each initiated projects to ascertain the protective scope of established human rights in respect of thoughts, emotions and other mental states, both now and in the future. For example, according to the Committee on Bioethics of the Council of Europe, there is a need to assess

> "whether these issues can be sufficiently addressed by the existing human rights framework or whether new human rights pertaining to cognitive liberty, mental privacy, and mental integrity and psychological continuity, need to be entertained in order to govern neurotechnologies. Alternatively, other flexible forms of good governance may be better suited for regulating neurotechnologies."[7]

Neurotechnologies have been scrutinized for raising human rights implications, in particular because neurotechnologies provide the capability to (1) *access* someone's mental states (2) *verify* subjective (or first-person) reports regarding the nature and content of those states, (3) *contest* first person authority regarding mental states by overriding such introspective reports, and (4) *control* decoded mental states by providing input behaviorally or through direct brain stimulation. But how exactly might neurotechnologies present human rights challenges?

Ethical challenges have emerged in all domains of application including the medical, the consumer domain, and the use of neurotechnologies by state actors such as in the military and criminal justice contexts.[8] Let us look at this latter domain in greater detail.

Scholarly consideration has been given, for example, to the use of brain-reading neurotechnology in the context of criminal justice.[9] The primary, though still not practically applicable, ethical concern in this domain is the possibility that in the course of the investigation of crimes, the police might employ some form of brain-reading neurotechnology to make inferences about the mental processes of suspects in order to advance their investigation. An even more speculative concern is that in the future neurotechnologies might be employed as a sentencing option, for example a closed-loop device could be used to monitor the brain of an offender and to intervene upon it in order to avert an angry outburst that might precipitate an offense.[10] Such scenarios give rise to questions of mental privacy, cognitive liberty and mental integrity and it is not hard to imagine examples from the other domains mentioned in the last paragraph that might generate disquiet.

In view of the aforementioned technological developments and emerging risk-assessment evaluations, questions arise as to whether existing human rights are sufficient to protect our brains and minds. While some authors have argued so, others advocate for the introduction of novel brain-specific rights, so called 'neurorights'.[11]

Whether neurorights are needed or not is a question currently being addressed by academics, as well as national legislators and intergovernmental organizations. For example, national legislators in Chile have been working on the implementation of neurorights into their national legal systems within the framework of a constitutional reform.[12] At the level of human rights, the United Nations, the Inter-American Juridical Committee, and the Council of Europe are exploring whether the established scope of existing rights and freedoms provides sustainable legal protection to our brains and minds in view of emerging neurotechnologies.[13] Recently, UNESCO, too, contributed to a report on the risks and challenges of neurotechnologies for human rights.[14]

While these policy initiatives are still in the making, central notions in this area such as 'mental privacy', 'mental integrity' or 'cognitive liberty' are not subject to consensus. Notably, there is a relative paucity of interdisciplinary theorizing and conceptualization, integrating perspectives from science and technology (neuroscience, neuroengineering and computer science), philosophy, ethics, legal philosophy and law.

This paper constitutes a concerted interdisciplinary effort by a group of scholars from different academic fields – inter alia, philosophy, biomedical ethics, law, neuroscience, cognitive science, medicine – in mapping the ethical and legal foundations of 'neurorights' with the purpose to facilitate international discussions and policy initiatives at various levels.[15]

To this end, we first briefly introduce the notion of neurotechnology and its current applications. We then map the current state of the ethical and legal debate on the relation between neurotechnologies and human rights and critically appraise the notion of 'neurorights'. In the second part, we will provide a philosophical and ethical analysis of three key notions in the neurorights debate, namely 'mental privacy', 'cognitive liberty' and 'mental integrity'. Finally, in the third part, we will examine the potential legal implications of this integrated conceptual understanding, particularly regarding current rights-based policy approaches to 'neuroprotection'. We will focus on three main areas of human rights significance: mental privacy, mental integrity, and cognitive liberty.[16]

## 2. Neurotechnologies and their applications

Currently, several types of neurotechnologies are under development. While there are many ways in which we can intervene in the brain through pharmaceuticals, psychotherapy and other means, the term 'neurotechnology' denotes devices that measure brain structure or function (particularly brain activity) or intervene into brain activity (e.g. through electrical stimulation). Typical examples of neurotechnology are brain-computer interfaces (BCIs), i.e. systems that measure and analyze brain activity to control an 'effector' (such as a robotic arm, or a software for text generation).

Three central types of neurotechnology are to be distinguished (see **Table 1**):

1. Devices that can monitor brain activity, i.e. neural correlates of mental states and behavior;
2. Devices that can intervene into brain activity, e.g. through electrical stimulation;
3. Bimodal devices that combine type 1 and 2.

Apart from their use in medicine, neurotechnologies can be used also in other domains such as the consumer market (e.g. BCIs for gaming, education, or meditation),[17] criminal justice (e.g., potentially as tools for lie detection),[18] and the military (e.g potentially for cognitive enhancement or covert brain-to-brain communication).[19] Examples of type 1 include passive brain computer interfaces for device control (with no stimulation capacity),[20] neuroprediction systems,[21] and fMRI-based techniques.[22] Examples of type 2 include invasive[23] and non-invasive[24] brain stimulation, for example to treat neurological or psychiatric disorders. An example of type 3 is closed-loop neuroadaptive brain stimulation via implants that monitor brain activity and deliver stimulation based on these measurements (thereby 'closing the loop') with the aim to modulate brain activity (e.g. in treating Parkinson disease).[25] While neurotechnologies today are mostly developed for use within individual brains, there are already proof-of-concept studies that demonstrate the possibility of brain-to-brain interaction mediated by neurotechnology.[26]

| Type of neurotechnology | Non-invasive | Invasive |
|---|---|---|
| **Type 1**: measuring brain structure or function | Extracranial Electroencephalography (EEG); (functional) magnetic resonance imaging ((f)MRI); Magnetoencephalography (MEG); Near-infrared spectroscopy (NIRS) | Intracranial EEG / electrocorticography (ECoG), i.e. surface grid electrodes for measuring brain activity from the brain surface; Stereo-Electroencephalography (sEEG), i.e. stereotactically implanted depth electrodes in specific brain regions |
| **Type 2**: intervening in brain | transcranial direct current | deep brain stimulation (DBS); |

| structure or function | stimulation (tDCS); (repetitive) transcranial magnetic stimulation ((r)TMS); transcranial focussed ultrasound (fUS) | sEEG, grid electrodes for epidural stimulation; vagal nerve stimulation (VNS) |
|---|---|---|
| **Type 3**: measuring and intervening in brain structure or function | Systems combining Type 1 and Type 2, e.g. a brain-computer interface (BCI) consisting of an extracranial EEG and extracranial stimulation, e.g. in a closed-loop setup | Systems combining Type 1 and Type 2, e.g. an implantable ECoG grid electrode that can both measure brain activity and deliver brain stimulation, e.g. in a closed-loop setup |

## 2. Neurotechnologies and human rights: the current state of the legal debate

The ethical and legal implications of neuroscience research, as well as the regulation and use of neurotechnologies, have received attention since the 1990s. This decade was often referred to as the 'decade of the brain'[27] as it witnessed a substantial boost in funding for neuroscience and neurotechnology research. This trend was subsequently amplified by the later emergence of large-scale research projects such as the EU's Human Brain Project and the US BRAIN Initiative in the 2000s and 2010s.[28] Since the inception of these large-scale projects, a crucial societal topic has been the legal protection of the human person against misuses of neurotechnology by governments or private actors (such as individual users, user groups but also companies that develop and/or manufacture neurotechnological devices).

It is beyond question that a set of established human rights applies to many conceivable scenarios of neurotechnological use. They include the rights to bodily integrity, privacy, personal identity, freedom of thought, and autonomy. However, a need for a widening of existing rights has been examined through discussions of the idea of "cognitive liberty" since the turn of the millennium.[29] This debate has recently received renewed interest under the heading of 'neurorights', an umbrella term used to encompass the set of rights that should guarantee adequate protection of the mind and brain of the human person. Although the terminology sometimes diverges, three neurorights families appear central to the present debate: a right to mental privacy, a right to mental integrity, and a right to cognitive liberty.[30]

From the current debate, it seems that legal protection of the mind might be pursued in the following ways: (1) under international and supranational human rights law[31]; (2) under domestic constitutions; (3) under ordinary domestic laws (including e.g. applied areas such as consumer protection law); (4) by some combination of 1, 2, 3 in a multi-level legal approach or framework. As previously mentioned, this paper focuses on human rights at the level of international and supranational law.

In the discussion on human rights protection of the brain and mind, three main positions can be distinguished.

    1. *Novel rights for specifically protecting the brain and mind ('neurorights') are necessary*. According to this position, the scope of protection of existing rights and freedoms is insufficient to offer

adequate legal protection from misuse of emerging neurotechnologies. At the time these rights were introduced, no one could have foreseen the possibilities neurotechnology offers today to monitor and intervene in our brains and, ultimately, in our mental states and behaviors. Hence, in the development of existing rights, the intricacies of game-changing neurotechnologies have not been taken into account. Even though it might be possible to broaden their protective scope in the course of their application, ultimately by court decisions, new neurorights must be introduced in order to fill the current gap in legal protection in a swift manner.[32]

2. *Adaptive interpretations and applications of existing rights are necessary, but novel neurorights are not.* Scholars defending this position agree with advocates of position 1 that the scope of protection offered by established rights and freedoms is insufficient to offer adequate protection with respect to emerging neurotechnologies. However, instead of advocating the introduction of novel rights and freedoms, they hold that it would be sufficient to update our interpretations of existing human rights.[33] Ultimately, the precise level of legal protection will depend on how we *specify* the protective scope of existing rights *through interpretation and application* and what kind of positive obligations can be derived from them for state actors to ensure their enjoyment in the context of emerging neurotechnologies. Current rights that are prima facie well-positioned to ensure legal protections for minds and brains include the right to privacy, the right to freedom of thought, and the right to mental integrity. Apart from human rights, derived (individual-subjective) rights should be taken into account as well, such as those defined on the national or supranational level, for example by the General Data Protection Regulation (GDPR) in the European legal context.[34]

3. *No novel rights, reforms or new interpretations are necessary.* The third position argues that the existing human rights suffice to offer effective legal protection from the misuse of emerging neurotechnologies. Therefore, novel rights and evolutionary interpretations of existing rights are both unnecessary. This type of argument, as of now, is not very prevalent in the current academic literature on the topic but, in the experience of the authors, is often invoked in discussions in legal and policy fora on neurorights.[35]

In discussing these positions, much emphasis is given to contingent technological considerations of the current capabilities of neurotechnologies to enable inferences on internal mental states from brain measurements or to interfere with brains and mental states. However, regardless of the contingent state of neurotechnology at any given time, the analytic question of whether we should develop specific protections for the brain and mind has critical importance.[36] Comparing and critically evaluating the various normative and conceptual stances at a high level is essential to advance the debate on 'neurorights', identify both conceptual divergences and areas of common ground as well as to provide policymakers with a conceptually solid foundation for present and future policy work in this area.

Interdisciplinary discussions on neurorights require clarity about rights. Moral values are often couched in the language of rights, since rights have become a key currency in which societal conflicts are addressed and resolved. But many do not demand legal rights in a strict sense. Rather, the ethical debate around neurorights wishes to affirm moral interests or values and many derive claims for stronger legal protection from these ethical considerations. It is important to bear in mind that legal rights are not merely the legal variations of moral claims or interests. They are technical entities, interwoven with the legal frameworks in which they are embedded, and interrelated to other rights and norms.[37]

A further important point about legal rights[38] is that they can be situated at different levels: in regulations or statutes of ordinary positive law, in constitutional or even international law. At these levels, legal rights are abstract and general, albeit to varying degrees, and apply to innumerable situations. Because of this, they often have to be more precisely defined by those applying the law and courts so that they can be applied in specific cases. In addition, rights at the latter two levels have been argued to be both more and less powerful than rights at the former levels. It should be noted, however, that it seems debatable how powerful international rights are because in many countries, international law generally has no effect in the courts unless a parliament has chosen to give effect to it through legislation.

In this paper, we will not take a position about the need for specific neurorights to protect the core subject matter identified in section 3. Rather, we aim to advance the debate by considering the extent to which mental integrity, mental privacy and cognitive liberty enjoy protection under the established framework of human rights law (section 4). This systematic analysis based on the normative foundations discussed in section 3, will be relevant to all three central positions regarding the development of specific neurorights on the international and regional level.

## 3. Conceptual and ethical foundations of mental integrity, mental privacy, and cognitive liberty

We recognize three core families of ethical and legal entitlements that can be construed as so-called neurorights. These are: mental integrity, mental privacy, and cognitive liberty. It should be noted that whereas mental integrity and mental privacy have been established as predominantly negative rights, as they are meant to ensure a *freedom from* external coercion or interference with agents' brains and minds, cognitive liberty, however, has been interpreted more broadly to include both negative freedom from coercion or interference and positive *freedom to* control one's own brain and mind.

### 3.1 Mental integrity

Conceptually, mental integrity can be understood as an analogue of the better-understood and more widely recognized notion of bodily integrity. From a moral perspective, in a minimalist conception, the right to bodily integrity protects against certain forms of interference with one's body. By analogy, the right to mental integrity, on a minimalist conception, would protect against certain forms of interference with one's mind.

We suggest that a moral right to mental integrity can coherently be endorsed, and distinguished from a moral right to bodily integrity, even if the mind is merely part of, or wholly resides in, the body. On these views, the right to mental integrity could be understood as a right over certain parts or functions of the body, i.e. the brain.[39]

As well as mirroring the right to bodily integrity in its content, the right to mental integrity might be thought to share its justification with the right to bodily integrity. The right to bodily integrity is often thought to be grounded in rights of self-ownership or personal sovereignty.[40] Since the mind is enabled by the body (the central nervous system in particular), and self and personhood are functions of the mind, it is plausible that if self-ownership and personal sovereignty ground a right to bodily integrity, they also ground a right to mental integrity.[41]

Some might argue that, given that the mind is part of and enabled by the body, and given that the body already enjoys the established protection of an integrity-right, a right to mental integrity would be superfluous. However, this objection does not take into account that the kinds of bodily interference that infringe the right to bodily integrity do not necessarily correspond to the kinds of mental interference that infringe a right to mental integrity. Consider, for example, so-called 'non-invasive' brain stimulation (NIBS), such as transcranial direct current stimulation (tDCS) and transcranial magnetic stimulation (TMS), which stimulate the brain via electrodes or electromagnets placed on the scalp. Nonconsensual NIBS could amount to a severe mental interference, and thus a serious violation of the right to mental integrity, even though, given the absence of significant changes in brain anatomy, they arguably involve no or only a minor interference with the body.

The idea of a right to mental integrity raises several questions. For example, which kinds of mental influence qualify as mental interference, and which mental interferences, precisely, infringe upon the right to mental integrity? Does the interference have to be harmful, for example, or at least to be significant in its effects? Given that in social dynamics every actor constantly influences other actors—including at the level of beliefs, desires and feelings—the right to mental integrity threatens to be very expansive. Therefore, it would need to be specified in such a way that distinguishes violations of mental integrity from innocuous and inevitable forms of mental influence, such as bona fide argument-based persuasion.[42]

### 3.2 Mental privacy

Mental privacy has long been considered an important feature of human individuality and freedom. Several features of mental privacy can be distinguished.[43] Mental states can be incommunicable or inaccessible to others in the sense that people can experience insurmountable difficulties in adequately expressing their thoughts or feelings. That is, there can be a felt difference between the report and the experience of that which is reported, in part due to the special access one has to one's own inner life. Mental states can also be taken to be unchallengeable. "That's how I feel it" is a statement that, in many cases, invokes an unassailable authority regarding the knowledge about one's mental life from the first-person perspective. Therefore, the potential implications of brain reading for a person's privacy and authority regarding their mental life are substantial.

First of all, the possibility that thoughts and feelings could be observed, not only indirectly through behavior but also through multimodal data analysis in which data about brain states play a substantial role, may have significant consequences for the individual, their social interactions, and political freedoms. Privacy as 'the right to be let alone'[44] can be considered as an important precondition for human freedom and dignity[45]. The decoding of brain data may, one day, reveal mental information such as someone's sexual preferences and political orientation, potentially leading to discrimination and prejudicial treatment. Knowing a person's preferences and emotions (e.g. fears) could enable another actor to control that person's behavior.[46] Obviously, human rights law recognizes a right to respect for private life, the protection of personal data, the right to freedom of thought, and the freedom to hold opinions. However, one could argue that there are important aspects regarding the relation of the right to mental privacy and freedom of thought that need to be considered from an ethical and legal perspective: for example, having free thoughts that are nevertheless monitored continuously could be seen as a violation of mental privacy under certain circumstances (e.g. in the absence of ongoing consent).

On the positive side, the possibility of effectively 'reading' the mind would be of a paramount importance in aiding patients who , for whatever reason, may have lost their capacity to communicate and/or move (e.g., patients with severe motor impairment or disorders of consciousness[47]) and helping them to protect their rights. For example, such technology can be used to allow people with communication disabilities to express their voluntary and informed consent.[48] Similarly, it can help people with physical disabilities to help themselves, or, in case of psychological problems, enable or allow themselves to be helped[49]. In cases where someone's self-understanding or self-control is diminished or failing, accessing brain states and making them available to the person in question through processes of neurofeedback[50] may lead to behavioral and cognitive improvements.

Second, being able to access the neural bases of mental states implies that one can evaluate the veracity of an individual's reports. This may have important consequences in legal contexts, e.g. in criminal law concerning the truthfulness of statements of suspects or witnesses (lie detection, brain fingerprinting)[51].

Third, by examining the brain one may attempt to override and correct an individual's reports about thoughts and feelings. One could advocate (or contest) for instance the use of brain reading in tort law for an objective assessment of the validity of claims about damage (e.g. pain) resulting from an incident.[52] In a psychotherapeutic context, one can also consider utilizing the more profound knowledge of brain states to aid patients[53].

When considering mental privacy in the face of neurotechnologies, it is useful to make some clarifications with respect to the different levels at which interference with the personal sphere of the individual can occur. This allows us to better qualify the kind, content, and ranking of rights that are relevant to potential violations of mental privacy.

It is often claimed that a great deal of personal information is disseminated by individuals either voluntarily (for example, through social media interactions) or unintentionally, through many electronic tracking and logging systems that we implicitly or explicitly consent to, are unaware of, or do not care about at all. However, it appears that it is difficult to propose a rigid application of mental privacy in a context in which information circulates in large quantities and at great speed. The type of information that can be collected thanks to digital profiling (either lawfully or unlawfully), makes it possible to track, analyse and predict many attitudes and behaviours of individuals, even those that are more sensitive, including those relating to as sexual orientation, political leaning, or health status.

It can therefore be inferred that the information residing in the individual's mind and brain is potentially more sensitive and subjectively relevant to the individual compared to non-mental information, as it is otherwise inaccessible to the direct knowledge of others. The mind and brain are thus, even in comparison with the difficulty of keeping other data confidential, the ultimate seat of personal information and the individual's refuge of privacy, to which a special value and, therefore, special protection is to be attributed.

### 3.3 Cognitive liberty

If the right to mental privacy may help protect the mind from external access and inspection, the principle of cognitive liberty has been invoked to protect mental states from external influence and interference. This human rightcandidate has been defined as a person's autonomous, unhindered control or mastery over their mind. For example, Bublitz has described cognitive liberty as a synonym for "mental self-determination."[54] In his account, this right comprises, among others, two fundamental and intimately related principles: (a) the right of individuals to freely use emerging neurotechnologies; (b) the protection of individuals from the coercive and unconsented use of such technologies. In other words, cognitive liberty is the principle that guarantees "the right to alter one's mental states with the help of neurotools as well as to refuse to do so".[55] A similar account is provided by Farahany, who defines cognitive liberty as "the right to self-determination over our brains and mental experiences".[56] She advances cognitive liberty as an updated concept of liberty for the digital age, rather than a "neurorights" concept limited to neurotechnologies. In this view, the right to cognitive liberty encompasses a broad spectrum of freedoms and rights such as the "freedom of thought and rumination, the right to self-access and self-alteration, and to consent to or refuse changes to our brains and our mental experiences" (p. 98). In Farahany's view, the right to cognitive liberty "is not absolute, but must be balanced against the societal costs it introduces". Ienca and Andorno have also recognized the dual nature of cognitive liberty by arguing that it constitutes a "complex right which involves the prerequisites of both negative and positive liberties in the sense of Isaiah Berlin (Berlin, 1969)"[57] In its negative sense, cognitive liberty describes the liberty of making choices about one's own cognitive domain in absence of external obstacles, barriers or prohibitions as well as of exercising one's own right to mental integrity in absence of external constraints or violations. In its positive sense, it involves the capability and right of acting in such a way as to take control of one's mental life. They also acknowledge the non-absolute nature of this right.

Many authors have emphasized the conceptual proximity between cognitive liberty and the right to freedom of thought. Some of them, such as Lavazza,[58] have argued that the existing right to freedom of thought is normatively well-suited to address all human rights challenges raised by neurotechnology and other emerging technologies. Other authors, such as Ienca,[59] have argued that the primary scope of the right to freedom of thought did not concern cognitive and affective processes (*forum internum*). In contrast it concerned societally-embedded entitlements such as freedom of conscience and religion. According to this view, although cognitive liberty and freedom of thought are intimately intertwined, they are better understood as distinct rights. Similarly, Farahany has argued that cognitive liberty is broader than freedom of thought because it also includes the rights to self-determination (self-access and self-alteration, and the right to consent to or refuse changes to our brains and our mental experiences), and mental privacy, as three overlapping but necessary components to cognitive liberty.[60]

## 4. Legal foundations of mental integrity, mental privacy, and cognitive liberty

It should be noted that the three notions described above are not to be seen as standalone normative principles. For example, in most cases, in order to influence someone's mental states it is necessary to have access to mental information in the first place. In those cases, mental privacy and cognitive liberty are both inevitably at stake. Similarly, unauthorized modifications of a person's mental states may qualify as violations not only of cognitive liberty but also of mental integrity. Therefore, the triadic classification above should be seen as purely taxonomical, not as indicative of (an absence of) relational dynamics.

Taxonomic considerations aside, one crucial question is determining how the protection of mental integrity, mental privacy, and cognitive liberty could be anchored in established human rights. Therefore, in the next section we will attempt to point to legal foundations of these prima facie rights and map their relationship to established human rights. Generally, we will take the right to mental integrity to refer to the protection from certain forms of unwanted or unwarranted interference with one's mind, the right to mental privacy to protect against certain forms of access to one's mind, and the right to cognitive liberty to protect one's mental self-determination.

**4.1 The right to mental integrity**

At present, a right to mental integrity has been recognised under different international and regional human rights instruments. At the international level, neither the Universal Declaration of Human Rights (UDHR) nor the International Covenant on Civil and Political Rights (ICCPR) guarantee an explicit right to mental integrity. However, Article 17 of the UN Convention on the Rights of Persons with Disabilities (CRPD) prescribes that "Every person with disabilities has a right to respect for his or her physical and mental integrity on an equal basis with others."

Within the Inter-American context, Article 5(1) of the American Convention on Human Rights (ACHR) states that "Every person has the right to have his physical, mental, and moral integrity respected." The different dimensions of integrity described by Article 5 ACHR are associated with health, understood in a wide sense that encompasses complete physical, mental and social well-being. However, in the Declaration of the Interamerican Juridical Committee on Neuroscience, Neurotechnologies and Human Rights some worries have been raised about the scope of this right. Although the enforceable contents of the right are clear within the medical context (e.g., the right to informed consent and to medical confidentiality), they are less clear for neurotechnologies with non-medical purposes.[61]

In the European context, a right to mental integrity has been recognised in the jurisprudence of the European Court of Human Rights (ECtHR), alongside a right to bodily integrity. Both are part of the right to respect for private life, guaranteed by Article 8 of the European Convention on Human Rights.[62] Sometimes, the ECtHR also refers to a right to 'moral' and 'psychological' integrity, but the case law suggests that mental, psychological, and moral integrity are interchangeable terms.[63] The recognition of a right to mental integrity by the ECtHR is also reflected in the Charter of Fundamental Rights of the European Union (CFR), aiming to provide comprehensive protection of the person, especially against novel technologies.[64] According to Article 3(1) CFR, "Everyone has the right to respect for his or her physical and mental integrity."

Although a right to mental integrity has an explicit basis in established human rights law, the exact scope and limits of this right have not yet been clarified.[65] Accordingly, its implications with respect to neurotechnology are still unclear.[66] Meanwhile, in the European context, some contours of the right to mental integrity have been sketched. This could provide helpful directions in the debate about neurotechnology and mental integrity. For example, case law on Article 8 ECHR illustrates that mental health is to be considered a crucial part of private life associated with mental integrity.[67] Furthermore, the right to mental integrity appears to cover complaints about bullying at school,[68] well-founded fear for physical abuse,[69] and loss of honor and reputation.[70] In addition, the ECtHR considers that, together with the notion of physical integrity, a person's mental integrity embraces multiple aspects of a person's identity, such as gender identification, sexual orientation, name, and elements relating to a person's right to image.[71]

Furthermore, the EU Network of Independent Experts on Fundamental Rights, set up by the European Commission, considers that the right to mental integrity pursuant to Article 3(1) CFR "is a fairly broad right".[72] It not only includes the prohibition of mental torture, inhuman and degrading treatment and punishment, but also covers a broad range of less serious forms of interference with a person's mind, which have traditionally been covered by the right to privacy. Examples that are provided concern mandatory treatment with psychoactive drugs, forced psychiatric interventions, strong noise, and 'brainwashing'.[73] Interferences with the mind through neurotechnology may well fit within this series of mind-altering interventions that are covered by the right to mental integrity in the meaning of Article 3 CFR.

Interestingly, apart from a right to mental integrity, changing people's minds (without consent) through means such as brainwashing is covered by the existing right to freedom of thought as well, at least to some extent.[74] The inner part of this right – the freedom to have, adopt, and change thoughts, conscience and religion – seeks at its most basic level "to prevent state indoctrination of individuals by permitting the holding, development, and refinement and ultimately change of personal thought, conscience and religion."[75] This internal dimension of freedom of thought is absolute. It protects freedom of thought unconditionally.[76] The scope of the right to freedom of thought is considered to be broad, as it not only covers the protection of religious thoughts and convictions.[77] In addition, it pertains to freedom of thought 'on all matters', encompassing a whole range of personal and collective convictions in the political, economic, social, scientific and intellectual spheres.[78]

Meanwhile, the exact scope of the right to freedom of thought is as-yet unclear. Some have defended a broad scope so as to cover a wide range of neurotechnologies that affect mental properties such as thoughts, memories and, possibly, emotions.[79] Others have suggested a somewhat restricted understanding,[80] which might not cover just any mind-altering technique and would thus, arguably, put more emphasis on the right to mental integrity to protect against unwanted mental interference.[81] How exactly the qualified right to mental integrity relates to the absolute right to freedom of thought, is however an open question.[82]

**4.2 The right to mental privacy**

Unlike the right to mental integrity, a right to mental privacy has not yet been recognised as a specific human right, at least not explicitly. However, the privacy of our thoughts, emotions and other mental states seems to gain at least some *implicit* legal protection under three distinctive human rights and freedoms. These are (1) the right to privacy, (2) the right to freedom of thought, and (3) the right to freedom of expression. Mental privacy as a specific right vis a vis neurotechnology is characterized differently by its kind and content and, consequently, is a potential candidate for a higher rank. This is justified by the fact that the tools capable of brain reading can access personal information (thoughts, judgements, desires, intentions) that the individual has never manifested and could never manifest externally.[83] This can actually happen today, for instance with communication neurotechnology based on detecting and interpreting neural signals to produce intelligible speech, writing or typing.[84] Let's consider current human rights protection of mental privacy in some more detail below.

*The right to privacy*

Article 12 UDHR prescribes that "No one shall be subjected to arbitrary interference with his privacy, family, home or correspondence, nor to attacks upon his honor and reputation. Everyone has the right to the protection of the law against such interference or attacks." A similar right is guaranteed by Article 17 ICCPR. According to the UN Human Rights Council (HRC), privacy can be considered as the presumption that individuals should have an area of autonomous development, interaction and liberty. People should have a 'private sphere' with or without interaction with others, free from State intervention as well as from excessive unsolicited interventions by other uninvited individuals.[85] Article 17 ICCPR pertains to all interferences with a person's privacy, regardless of whether they emerge from State officials or from natural or legal persons. States must adopt legislative and other measures to give effect to the prohibition against privacy interferences and to the protection of Article 17 ICCPR[86] In the age of technology and digital environments, the HRC considers *informational privacy* of particular importance, covering information that exists or can be derived about a person and her life and to decisions based on that information.[87]

The scope of the international human right to privacy is broad and able to adapt to modern means of privacy interference, such as through neurotechnology. For example, today, it already covers the protection of metadata, as such data, when analysed and aggregated, "may give an insight into an individual's behaviour, social relationship, private preference and identity".[88] Furthermore, the HRC refers to data-driven technologies, which "increasingly enable States and business enterprises to obtain fine-grained information about people's lives, make inferences about their physical and mental characteristics and create detailed personality profiles."[89]

Within the Interamerican context, Article 11 of the ACHR recognizes a right to privacy. Mirroring the UDHR and the ICCPR, it articulates a right to protection from arbitrary or abusive interferences with private life, family, home or correspondence and unlawful attacks on honor or reputation. Interestingly, a detailed account of informational privacy has also been provided by the Interamerican Juridical Committee (CJI). The 2021 Updated Principles of the Interamerican Juridical Committee on Privacy and Personal Data Protection presents a specific set of rules for data protection. These include transparency, consent, justification of the relevance and necessity of data processing, the restriction of data retention and processing, data confidentiality, security and accuracy, and the facilitation to data owners of access, rectification, erasure and portability of data. Regarding mental privacy, it is worth mentioning the definition of Sensitive Personal Data provided by these Updated Principles, which affirms that some types of personal data "are especially likely to cause material harm to individuals if misused [and] [d]ata controllers should adopt reinforced privacy and security measures that are commensurate with the sensitivity of the data and its capacity to harm the data subjects."

The recent Declaration of the Interamerican Juridical Committee on Neuroscience, Neurotechnologies and Human Rights, specifically refers to this principle in its justification of the idea that neural data protection may require updating traditional formulas for the protection of privacy, in other to prevent different ways in which its collection, processing and application could affect the autonomy and personality of individuals.

The Declaration suggests that the special protection required by neural data is grounded in how the revealing of these data could affect a person's identity and autonomy, highlighting that neural data could be used to manipulate mental processes and, ultimately, behavior: the "development of neurotechnologies can lead to the conditioning of personality and the loss of autonomy of individuals, and in this context one of the most pressing concerns has to do with the malicious behavior of those who access the people's brain activity data in order to penetrate their minds, condition them, or take advantage of such knowledge."

In the European context, a right to respect for private life has been recognized by Article 8 ECHR. When interpreting this right, the ECtHR considers the notion of personal autonomy as an important principle.[90] It argues that the protection of personal data is of fundamental importance to the enjoyment of the right to respect for private life pursuant to Article 8 ECHR. Accordingly, the domestic laws "must afford appropriate safeguards to prevent any such use of personal data as may be inconsistent with the guarantees of this provision. Article 8 ECHR thus provides for the right to a form of *informational self-determination*, allowing individuals to rely on their right to privacy as regards data which, albeit neutral, are collected, processed and disseminated collectively and in such a form or manner that their Article 8 rights may be engaged."[91] In considering whether personal data relates to 'private life' in the meaning of Article 8 ECHR, and whether the collection, storage, or use of that information will infringe this right, the ECtHR takes account of the specific context in which the data have been recorded and retained, the nature of the data, the way in which they are used and processed, the results that may be obtained, and, sometimes, the reasonable expectations of a person's privacy.[92]

Similar to the ICCPR, Article 8 ECHR has the ability to adapt to current developments in society, including developments in technology and bioethics.[93] Furthermore, in the European context, Article 7

CFR guarantees the right to respect for one's private life and Article 8 CFR covers the protection of personal data, detailed by the General Data Protection Regulation (GDPR) in EU secondary law. In this regard, personal data has been defined as "information relating to an identified or identifiable natural person ('data subject'); an identifiable natural person is one who can be identified, directly or indirectly, in particular by reference to an identifier such as a name, an identification number, location data, an online identifier or to one or more factors specific to the physical, physiological, genetic, mental, economic, cultural or social identity of that natural person."[94]

In sum, it is clear that people's privacy, and more specifically, their personal data, deserves considerable legal protection. It is also clear that brain-derived data will in most cases qualify as protected 'personal data', as they often relate to an identified or identifiable individual.[95]

*The right to freedom of thought*

Another human right that is relevant to the protection of mental privacy is the right to freedom of thought (briefly discussed in section 4.1).[96] The absolute, inner part of this right guarantees, amongst other things, that no one can be compelled to *reveal* one's thoughts or adherence to a religion.[97] According to the UN Special Rapporteur on Freedom of Religion or Belief, this implies that mental privacy "is a core attribute of freedom of thought. The right not to reveal one's thoughts against one's will arguably includes "the right to remain silent", without explaining such silence."[98]

Similar to the right to privacy, it is clear that established human rights law guarantees robust legal protection to the freedom of thought. Meanwhile, the extent to which the right to freedom of thought protects brain-derived data that enable to draw inferences about a variety of mental properties, is yet an open question. Much of the answer to this question will depend on how one understands the right to freedom of thought, having either a broad or a narrow scope.[99] A further question would be how the protection of mental privacy through the *absolute* right to freedom of thought would relate to the typically *qualified* protection of privacy rights, allowing for exceptions in certain situations . If mental privacy would be covered by both rights, there is a need to develop a theory that clarifies when information about the mind deserves absolute legal protection of the right to freedom of thought, over and above the qualified protection of the general right to privacy.[100]

*The right to freedom of expression*

Interestingly, under the ACHR, Article 13 assures a right to freedom of thought in conjunction with the freedom of expression: "Everyone has the right to freedom of thought and expression. This right includes freedom to seek, receive, and impart information and ideas of all kinds, regardless of frontiers, either orally, in writing, in print, in the form of art, or through any other medium of one's choice." Although formulations and levels of legal protection vary across different human rights instruments, the freedom of expression receives considerable protection under established human rights law, *inter alia* by 19 ICCPR, 13 ACHR, 10 ECHR, and Article 11 CFR. The scope of this right is conceived to be 'extremely broad'.[101] It extends to all forms of expression that impart or convey opinions, ideas or information of any kind. Moreover, it not only protects the *substance* of ideas and information that are expressed, but also covers the *means* by which they are manifested, transmitted, and received, ranging from speech, sign language and non-verbal expression through images and objects, to film, radio and internet-based modes of expression.[102]

The relevance of the freedom of expression to the protection of mental privacy lies in the fact that this freedom implies a negative aspect; a freedom *not* to express oneself. As the General Comment to Article 19 ICCPR puts it: "Freedom to express one's opinion necessarily includes freedom not to express one's opinion".[103] Regarding Article 10 ECHR, Harris et al. signal that "Albeit sparse in the case law, Article 10 guarantees to some extent the negative aspect of freedom of expression, namely the right not to be compelled to express oneself. One notable example is the right to remain silent."[104]

This negative freedom has the potential to embrace the liberty not to express or transmit by any means – such as through BCIs or other neurotechnologies – brain-related information that enables the drawing of inferences about a variety of mental states, irrespective of their content.[105] However, both in theory and in practice, the freedom of non-expression has only received little attention so far. Although it seems clear that the scope of this right is broad, its exact meaning and normative implications are to a large extent unexplored. This lack of well-defined doctrine entails uncertainties regarding the exact kind and level of legal protection the right to freedom of non-expression would offer to the notion of mental privacy. At the same time, the undeveloped nature of the freedom of non-expression can also offer opportunities. The precise scope and implications of this right are yet to be defined, as are the weight and importance of all kinds of competing interests in the balancing of this right. In developing a legal doctrine on the freedom of non-expression, scholars, judges, legislators, and policymakers can take into account the particularities neurotechnologies raise to the personal interest of mental privacy.[106]

To summarize, the notion of mental privacy receives some protection in the established systems of human rights, most notably by the right to privacy, the freedom of expression and the freedom of thought. Thus, the general framework to protect mental privacy is provided by established human rights. Now there is a need to specify this framework and to clarify how the different rights within that framework complement and relate to each other.

### 4.3 The right to cognitive liberty

At present, the legal protection of cognitive liberty has no explicit basis in established human rights law. However, it has been argued that the right to cognitive liberty could be seen as an extension or conceptual update of the existing right to freedom of thought (discussed in sections 4.1 and 4.2).[107] According to Farahany, cognitive liberty encompasses freedom of thought and rumination, the right to self-access and self-alteration, and to consent to or refuse changes to our brains and our mental experiences. Those concepts make up the "bundle of rights" that make up cognitive liberty, which translate to human rights to freedom of thought, self-determination, and mental privacy. She advocated cognitive liberty be recognized as a new human right, directing the updating of the three existing human rights the "bundle" implicates – freedom of thought, the collective right to self-determination, and privacy.[108] Farahany suggests cognitive liberty as a whole is to be balanced against other societal interests, while certain aspects of the bundle of rights –namely freedom of thought– are absolute.

Sometimes, the right to cognitive liberty is also referred to as a right to 'mental self-determination'.[109] That is, a right to control over the content of our own mental lives; a right to self-determine what is in and on your mind. Regarding the European context, it seems plausible that a right to mental self-determination could receive increased support in the near future – not necessarily as an extension or update of the absolute freedom of thought, but, possibly, also as a specification of the qualified right to respect for private life.[110] For example, the Committee on Bioethics of the Council of Europe writes in its Strategic Action Plan on Human Rights and Technologies in Biomedicine (2020-2025):

> "Technological developments in the field of biomedicine create new possibilities for intervention in individual behavior. For instance, certain technologies raise the prospect of increased understanding, monitoring, and control of the human brain, while other developments allow for the permanent health monitoring of individuals. These developments raise novel questions relating to autonomy, privacy, and even freedom of thought. (…) In the light of these developments, the third pillar of the Strategic Action Plan addresses concerns for physical and mental integrity. Guaranteeing respect for a person's integrity in the sphere of biomedicine is one of the central tenets of the Oviedo Convention. *This is understood as the ability of individuals to exercise control over what happens to them with regard to, inter alia, their body, their mental state, and the related personal data*."[111]

According to the Committee, the notion of 'mental integrity' – protected as part of private life under Article 8 ECHR – embraces the ability to control our own mental states. Within that approach, the right to mental integrity comes down to the freedom to exercise control over what is in and on our minds. In other words, in this conception, the right to mental integrity could comprise a right to mental self-determination; a right to cognitive liberty.[112]

Whether the ECtHR would be inclined to follow a similar, broad interpretation of the right to respect for private life is an open question. To date, it has accepted 'private life' to cover a right to *mental integrity*,[113] a generic right to *self-determination*,[114] as well as a specified right to *informational self-determination*.[115] Perhaps, the explicit recognition of a right to *mental self-determination*[116] may well fit within the Court's conception of the right to respect for private life under Article 8 ECHR.[117]

## 5. Concluding remarks

Advances in neurotechnology and artificial intelligence (AI) are not only challenging traditional boundaries of our brains and mental lives, they also challenge our traditional ways of thinking about human rights. More specifically, as our analysis has shown, there is agreement that mental privacy, mental integrity and cognitive liberty are discernible notions of moral concern that need to be considered in legal response to advances in neurotechnology. Both ethical and legal scholarship have highlighted the relevance of these concepts.

However, there are substantial differences in terms of how the philosophical and ethical foundations of these 'neurorights' are understood by scholars within and across academic fields. Since these notions can be *conceptualized differently* in terms of their philosophical and ethical foundations, it is debatable to what extent it is desirable and necessary to translate and condense them into specific legal rights at the international level and to integrate them into the existing human rights system. Therefore, to facilitate international debates on human rights protection of the mind, the ethical and philosophical understandings of 'neurorights' need to be, at least, made transparent and explicit and, ideally, be harmonized across different fields and perspectives.

To this end, our paper – a concerted interdisciplinary effort by scholars from a variety of academic fields – mapped the ethical and legal foundations of central 'neurorights'. The *minimalist conceptual understandings* of the key notions that have been provided could, we hope, facilitate further discussion and help develop legal interpretation and policy initiatives in this area. Such work is necessary to enable the implementation of legal protections not only at the international human rights level, but also in regional or national contexts.

**Acknowledgments**

AWP is funded by FONDECYT INICIACIÓN 11220327, *Funding Agency*: Agencia Nacional de Investigación y Desarrollo (Chile).

GM and SL are funded by the Dutch Research Council (NWO Vici grant VI.C.201.067).

The work of PK was partly funded by a grant by the Klaus Tschira Foundation (grant no. 00.001.2019).

**Conflict of interest**

No conflicts of interest.

**References**

---

[1] e.g. Ienca M, Andorno R. Towards new human rights in the age of neuroscience and neurotechnology, *Life Sciences, Society and Policy* 2017;13(1):5; Sommaggio P, Mazzocca M, Gerola A, Ferro F. Cognitive liberty. A first step towards a human neuro-rights declaration, *BioLaw Journal - Rivista di BioDiritto*. 2017:27–45; McCarthy-Jones S. The Autonomous Mind: The Right to Freedom of Thought in the Twenty-First Century, *Frontiers in Artificial Intelligence* 2019;2; Bublitz J-C. The Nascent Right to Psychological Integrity and Mental Self-Determination, in A. von Arnauld, K. von der Decken, M. Susi, eds. *The Cambridge Handbook of New Human Rights: Recognition, Novelty, Rhetoric*. Cambridge: Cambridge University Press; 2020:387–403; Michalowski S. Critical Reflections on the Need for a Right to Mental Self-Determination, in A. von Arnauld, K.

von der Decken, M. Susi, eds. *The Cambridge Handbook of New Human Rights: Recognition, Novelty, Rhetoric*. Cambridge: Cambridge University Press; 2020:404–12; Alegre S. Rethinking Freedom of Thought for the 21st Century, *European Human Rights Law Review* 2017;(3); Ligthart S. Freedom of thought in Europe: do advances in 'brain-reading' technology call for revision?, *Journal of Law and the Biosciences* 2020;7(1); Ligthart S, Douglas T, Bublitz C, Kooijmans T, Meynen G. Forensic Brain-Reading and Mental Privacy in European Human Rights Law: Foundations and Challenges, *Neuroethics* 2021;14(2):191–203.

[2] Report on Respecting, Protecting and Fulfilling the Right to Freedom of Thought, to the 76th Session of the General Assembly, October 2021; United Nations, Our Common Agenda – Report of the Secretary-General, New York 2021, par. 35.

[3] Declaration of the Interamerican Juridical Committee on Neuroscience, Neurotechnologies and Human Rights: New Legal Challenges for the Americas, CJI/DEC. 01 (XCIX-O/21, August 11, 2021)

[4] Committee on Bioethics of the Council of Europe, *Strategic Action Plan on Human Rights and Technologies in Biomedicine (2020-2025)*, Adopted by the Committee on Bioethics (DH-BIO) at its 16th meeting (19-21 November 2019).

[5] Report of the International Bioethics Committee of UNESCO, *Ethical Issues of Neurotechnology*, SHS/BIO/IBC28/2021/3Rev., 15 December 2021. See also M.S. Navarro et al, *The Risks and Challenges of Neurotechnologies for Human Rights* 2023.

[6] OECD, *Recommendation on Responsible Innovation in Neurotechnology*, Adopted by the OECD Council on 11 December 2019.

[7] See note 4, Committee on Bioethics of the Council of Europe 2019.

[8] Yuste R, Goering S, Arcas BA y, Bi G, Carmena JM, Carter A et al. Four ethical priorities for neurotechnologies and AI, *Nature* 2017;551(7679):159–63; Goering S, Klein E, Specker Sullivan L, Wexler A, Agüera y Arcas B, Bi G, et al. Recommendations for Responsible Development and Application of Neurotechnologies, *Neuroethics* 2021;14(3):365–86.

[9] Seaman JA. Your Brain on Lies: Deception Detection in Court, *The Routledge Handbook of Neuroethics*. Routledge; 2017; Farahany N. Searching Secrets, *University of Pennsylvania Law Review* 2012;160(5):1239; Farahany N. Incriminating Thoughts, *Stanford Law Review* 2012;64:351–408; *Neurolaw: Advances in neuroscience, justice, and security*, Palgrave Macmillan; 2021; McCay, A. Neurotechnology, law and the legal profession, Horizon Report for The Law Society (August 2022); Jotterand F. Punishment, Responsibility, and Brain Interventions, in F. Jotterand, ed. *The Unfit Brain and the Limits of Moral Bioenhancement*. Singapore: Springer; 2022:171–92.

[10] Gilbert F, Dodds S. Is There Anything Wrong With Using AI Implantable Brain Devices to Prevent Convicted Offenders from Reoffending?, in N. A. Vincent, T. Nadelhoffer, A. McCay, eds. *Neurointerventions and the Law: Regulating Human Mental Capacity*. Oxford University Press; 2020:0; Birks D, Douglas T, Birks D, Douglas T (eds.). *Treatment for Crime: Philosophical Essays on Neurointerventions in Criminal Justice*, Oxford University Press: Oxford, New York; 2018; See note 9, McCay 2022.

[11] See note 1, Ienca, Andorno 2017, at 5; J. Genser, S. Hermann & R. Yuste, *International Human Rights Protection Gaps in the Age of Neurotechnology*, report of the NeuroRights Foundation (May 2022).

[12] For some critical notes, see Zúñiga-Fajuri A, Miranda LV, Miralles DZ, Venegas RS. Chapter Seven - Neurorights in Chile: Between neuroscience and legal science, in M. Hevia, ed. *Developments in Neuroethics and Bioethics*. Academic Press; 2021:165–79; Fins JJ. The Unintended Consequences of Chile's Neurorights Constitutional Reform: Moving beyond Negative Rights to Capabilities, *Neuroethics* 2022;15(3):26.

[13] Report on Respecting, Protecting and Fulfilling the Right to Freedom of Thought, to the 76th Session of the General Assembly expected in July 2021; Declaration of the Interamerican Juridical Committee on Neuroscience, Neurotechnologies and Human Rights: New Legal Challenges for the Americas, CJI/DEC. 01 (XCIX-O/21, August 11, 2021; Ienca M. *Common human rights challenges raised by different applications of neurotechnologies in the biomedical field*, 2021; See note 4, Committee on Bioethics of the Council of Europe 2019.

[14] See note 5, Sosa Navarro 2022.

[15] As is often the case in interdisciplinary work, not all authors that contributed to the discussions and the resulting paper agree with every point made in the paper. We have made substantial efforts in harmonizing views and interpretations but also want to acknowledge the reality of "reasonable disagreement".

[16] It might be argued that drawing a distinction between bodily and mental integrity implicitly endorses a form of dualism between the body (including the brain) and the mind. To avoid this risk, some have proposed recognising a single right to "identity integrity". However, since the rights to bodily and mental integrity are widely used in the academic literature and acknowledged in the law, such as by the EU Charter of Fundamental Rights, we will stick to this terminology in this paper. In adopting this terminology, we do not mean to endorse dualism.

[17] Ienca M, Haselager P, Emanuel EJ. Brain leaks and consumer neurotechnology, *Nature Biotechnology* 2018;36(9):805–10; Williamson B. Brain Data: Scanning, Scraping and Sculpting the Plastic Learning Brain Through Neurotechnology, *Postdigital Science and Education* 2019;1(1):65–86; Kellmeyer P. Big Brain Data:
20

On the Responsible Use of Brain Data from Clinical and Consumer-Directed Neurotechnological Devices, *Neuroethics* 2021;14(1):83–98.

[18] Ligthart S, Toor D van, Kooijmans T, Douglas T, Meynen G (eds.), *Neurolaw: advances in neuroscience, justice & security*, Palgrave Macmillan: Cham, Switzerland; 2021. See note 10, Birks et al. 2018.

[19] Ienca M, Jotterand F, Elger BS. From Healthcare to Warfare and Reverse: How Should We Regulate Dual-Use Neurotechnology?, *Neuron* 2018;97(2):269–74.

[20] Roc A, Pillette L, Mladenovic J, Benaroch C, N'Kaoua B, Jeunet C et al. A review of user training methods in brain computer interfaces based on mental tasks, *Journal of Neural Engineering* 2021;18(1);. Krol LR, Haselager P & Zander TO. Cognitive and affective probing: a tutorial and review of active learning for neuroadaptive technology. Journal of Neural Engineering 2020;17:012001.

[21] Delfin C, Krona H, Andiné P, Ryding E, Wallinius M, Hofvander B. Prediction of recidivism in a long-term follow-up of forensic psychiatric patients: Incremental effects of neuroimaging data, *PLOS ONE* 2019;14(5):e0217127; Aharoni E, Vincent GM, Harenski CL, Calhoun VD, Sinnott-Armstrong W, Gazzaniga MS et al. Neuroprediction of future rearrest, *Proceedings of the National Academy of Sciences of the United States of America* 2013;110(15):6223–28.

[22] Farah MJ, Hutchinson JB, Phelps EA, Wagner AD. Functional MRI-based lie detection: scientific and societal challenges, *Nature Reviews. Neuroscience* 2014;15(2):123–31.

[23] Krauss JK, Lipsman N, Aziz T, Boutet A, Brown P, Chang JW, et al. Technology of deep brain stimulation: current status and future directions, *Nature Reviews Neurology* 2021;17(2):75–87.

[24] De Risio L, Borgi M, Pettorruso M, Miuli A, Ottomana AM, Sociali A, et al. Recovering from depression with repetitive transcranial magnetic stimulation (rTMS): a systematic review and meta-analysis of preclinical studies, *Translational Psychiatry* 2020;10(1):1–19.

[25] Bouthour W, Mégevand P, Donoghue J, Lüscher C, Birbaumer N, Krack P. Biomarkers for closed-loop deep brain stimulation in Parkinson disease and beyond, *Nature Reviews Neurology* 2019;15(6):343–52.; 1. Holmen SJ & Ryberg J. Interventionist Advisory Brain Devices, Aggression, and Crime Prevention. Journal of Cognition and Neuroethics 2021;8:1–22.

[26] Mashat MEM, Li G, Zhang D. Human-to-human closed-loop control based on brain-to-brain interface and muscle-to-muscle interface, *Scientific Reports* 2017;7(1):11001. For an overview see also: Kellmeyer P. Artificial Intelligence in Basic and Clinical Neuroscience: Opportunities and Ethical Challenges. Neuroforum 2019;25:241–250.

[27] Jones EG, Mendell LM. Assessing the Decade of the Brain, *Science* 1999;284(5415):739–739.

[28] Human Brain Project: https://www.humanbrainproject.eu/en/; NIH BRAIN Initative: https://braininitiative.nih.gov/

[29] The term was coined in a series of essays by Richard G Boire, On cognitive Liberty, in the Journal of Cognitive Liberties (http://www.cognitiveliberty.org/ccle1/1jcl/1jcl ).

[30] See note 132, Ienca 2021.

[31] In our discussion here, we will refer to specific 'neurorights' as 'human rights' if they are conceptualized (or discussed) within an international and universal context (e.g. in discussions at the level of the UN).

[32] Kellmeyer P. 'Neurorights': A Human Rights–Based Approach for Governing Neurotechnologies, in O. Mueller, P. Kellmeyer, S. Voeneky, W. Burgard, eds. *The Cambridge Handbook of Responsible Artificial Intelligence: Interdisciplinary Perspectives*. Cambridge: Cambridge University Press; 2022:412–26; See note 1, Ienca and Andorno 2017; See note 1 Sommaggio et al. 2017; See note 11, NeuroRights Foundation 2022.

[33] Bublitz JC. Novel Neurorights: From Nonsense to Substance, *Neuroethics* 2022;15(1):7; Bublitz JC. Freedom of Thought in the Age of Neuroscience: A Plea and a Proposal for the Renaissance of a Forgotten Fundamental Right, *ARSP: Archiv für Rechts- und Sozialphilosophie / Archives for Philosophy of Law and Social Philosophy* 2014;100(1):1–25; Ligthart S. *Coercive Brain-Reading in Criminal Justice: An Analysis of European Human Rights Law*, Cambridge University Press: Cambridge; 2022; Hertz N. Neurorights – Do we Need New Human Rights? A Reconsideration of the Right to Freedom of Thought, *Neuroethics* 2022;16(1):5; See note 1, McCarthy-Jones 2019 and Alegre 2017.

[34] Rainey S, McGillivray K, Akintoye S, Fothergill T, Bublitz C, Stahl B. Is the European Data Protection Regulation sufficient to deal with emerging data concerns relating to neurotechnology?, *Journal of Law and the Biosciences* 2020;7(1); See note 1, Ligthart et al. 2021.

[35] See note 1, Michalowski 2020; See note 1, Ligthart 2020; See note 122, Zúñiga-Fajuri et al. 2021.

[36] Douglas T, Forsberg L. Three Rationales for a Legal Right to Mental Integrity, in S. Ligthart, D. van Toor, T. Kooijmans, T. Douglas, G. Meynen, eds. *Neurolaw: Advances in Neuroscience, Justice & Security*. Cham: Springer International Publishing; 2021:179–201; Lavazza A. Freedom of Thought and Mental Integrity: The Moral Requirements for Any Neural Prosthesis, *Frontiers in Neuroscience* 2018;12; See note 8, Goering et al. 2021; Mecacci G & Haselager P. Identifying Criteria for the Evaluation of the Implications of Brain Reading for Mental Privacy. Science and Engineering Ethics 2019;25:443–461. Germani F, Kellmeyer P, Wäscher S & Biller-

[83] See DARPA's recent project on Neural Evidence Aggregation Tool (NEAT, https://www.darpa.mil/news-events/2022-03-02). "NEAT aims to develop a new cognitive science tool that identifies people at risk of suicide by using preconscious brain signals rather than asking questions and waiting for consciously filtered responses." See also: 1. Haselager P, Mecacci G & Wolkenstein A. in: *Clinical Neurotechnology meets Artificial Intelligence: Philosophical, Ethical, Legal and Social Implications* (eds. Friedrich, O., Wolkenstein, A., Bublitz, C., Jox, R. J. & Racine, E.) Cham: Springer International Publishing; 2021.55–68. doi:10.1007/978-3-030-64590-8_5;

[84] Chandler JA, Van der Loos KI, Boehnke SE, Beaudry JS, Buchman DZ, Illes J. Building communication neurotechnology for high stakes communications, *Nature Reviews Neuroscience* 2021;22(10):587–88; Maslen H, Rainey S. 'A Steadying Hand': Ascribing Speech Acts to Users of Predictive Speech Assistive Technologies, *Journal of Law and Medicine* 2018;26(1):44–53.

[85] UN Human Rights Council, The right to privacy in the digital age, A/HRC/39/29, 3 August 2018, par. 5.

[86] CCPR General Comment No. 16: Article 17 (Right to Privacy), par. 1.

[87] UN Human Rights Council, The right to privacy in the digital age, A/HRC/39/29, 3 August 2018, par. 5.

[88] UN Human Rights Council, The right to privacy in the digital age, A/HRC/39/29, 3 August 2018, par. 5.

[89] UN Human Rights Council, The right to privacy in the digital age, A/HRC/39/29, 3 August 2018, par. 15.

[90] ECtHR (GC) 5 September 2017, appl.no. 61496/08 (*Bărbulescu/Romania*), § 70.

[91] ECtHR (GC) 27 June 2017, appl.no. 931/13 (*Satakunnan Markkinapörssi Oy and Satamedia Oy/Finland*), § 137 (emphasis added).

[92] ECtHR (GC) 4 December 2008, appl.nos. 30562/04 and 30566/04 (*S. & Marper/UK*), § 67; ECtHR 13 February 2020, appl.no. 45245/15 (*Gaughran/UK*), § 70. See note 63, de Vries 2018, at 673.

[93] Council of Europe, *The European Convention on Human Rights: A living instrument*, Strasbourg 2020, p. 7.

[94] Article 4(1) GDPR.

[95] See note 4, Rainey et al. 2020; See note 1, Ienca, Andorno 2017.

[96] See note 33, Bublitz 2014; See note 1, Ligthart et al. 2021.

[97] General Comment No. 22: The right to freedom of thought, conscience and religion (Art. 18) CCPR/C/21/Rev.1/Add.4, par. 3; See note 74, Vermeulen 2006.

[98] Shaheed, UN Special Rapporteur on Freedom of Religion or Belief, Report on the Freedom of Thought, 5 October 2021, A/76/380, at 94, par. 26.

[99] See note 3333, Bublitz 2014; See note 1, Alegre 2017; See note 1, McCarthy-Jones 2019; See note 1, Ligthart 2020.

[100] See note 83, Ligthart, Bublitz, Forsberg, Douglas and Meynen 2022.

[101] Rainey B, Wicks E, Ovey C. *Jacobs, White and Ovey: the European Convention on Human Rights*, Eighth edition ed. Oxford University Press: Oxford, United Kingdom; 2020; See note 78, Harris 2018.

[102] General comment No. 34 Article 19: Freedoms of opinion and expression, CCPR/C/GC/34, par. 11-12; ECtHR (GC) 15 December 2005, appl.no 73797/01 (*Kyprianou/Cyprus*), § 174; Interamerican Commission on Human Rights, *Declaration of Principles on Freedom of Expression*, principle 2; Grossman C. Freedom Of Expression In The Inter-American System For The Protection Of Human Rights, *ILSA Journal of International & Comparative Law* 2001;7(3):619–47..

[103] General comment No. 34 Article 19: Freedoms of opinion and expression, CCPR/C/GC/34, par. 10.

[104] See note 78, Harris 2018, at 595. See e.g. EComHR 7 April 1994, appl.no. 20871/92 (*Strohal/Austria*); ECtHR (GC) 3 April 2012, appl.no. 41723/06 (*Gillberg/Sweden*), § 86; ECtHR 23 October 2018, appl.no. 26892/12 (*Wanner/Germany*), § 39-42. On an important note: This suggests that the right to silence has been protected by the ECtHR. The response of the ECtHR to English attacks on the right to silence suggest otherwise. One can remain silent, but adverse inferences can be drawn from the person's silence which does not amount to much of a protection of the right to silence. In the future we might expect the ECtHR to extend its approach by saying a person can refuse brain-based lie detection that the State wants to employ but the person does so, adverse inferences can be drawn from the refusal.

[105] See note 3333, Ligthart 2022; See note 1, Ligthart 2020.

[106] See note 3333, Ligthart 2022.

[107] Sententia W. Neuroethical considerations: cognitive liberty and converging technologies for improving human cognition, *Annals of the New York Academy of Sciences* 2004;1013:221–28; Bublitz C, Cognitive Liberty or the International Human Right to Freedom of Thought, in J. Clausen, N. Levy, eds. *Handbook of Neuroethics*. Dordrecht: Springer Netherlands; 2015:1309–33.

[108] See note 56, Farahany 2019 and Farahany 2023.

[109] See note 1, Bublitz 2020; See note 1, Ienca, Andorno 2017; See note 41, Bublitz, Merkel 2014.

[110] Ligthart S, Kooijmans T, Douglas T, Meynen G. Closed-Loop Brain Devices in Offender Rehabilitation: Autonomy, Human Rights, and Accountability, *Cambridge Quarterly of Healthcare Ethics* 2021;30(4):669–80. Kellmeyer P, Cochrane T, Müller O, Mitchell C, Ball T, Fins JJ, et al. The Effects of Closed-Loop Medical Devices on the Autonomy and Accountability of Persons and Systems. Cambridge quarterly of healthcare ethics 2016;25:623–633.